\documentclass[11pt]{article}
\usepackage{amsfonts,bm,bbm}
\usepackage{mathrsfs}
\usepackage{amsmath}
\usepackage{cite}
\usepackage{graphicx}
\usepackage{rotating}
\usepackage{hyperref}
\usepackage{verbatim}
\usepackage{color}

\csname @addtoreset\endcsname{equation}{section}
\newcommand{\beq}{\begin{equation}}
\newcommand{\eeq}{\end{equation}}
\newcommand{\bea}{\begin{eqnarray}}
\newcommand{\eea}{\end{eqnarray}}
\newcommand{\ba}{\begin{array}}
\newcommand{\ea}{\end{array}}
\newcommand{\bit}{\begin{itemize}}
\newcommand{\eit}{\end{itemize}}

\newcommand{\complesso}{{\ \hbox{{\rm I}\kern-.6em\hbox{\bf C}}}}
\newcommand{\reale}{{\hbox{{\rm I}\kern-.2em\hbox{\rm R}}}}

\newcommand{\1}{ \,  \raisebox{+0.14em}{{\hbox{{\rm \scriptsize ]}} \raisebox{-0.2em}{\kern-.8em\hbox{1}}}} \, }

\begin{document}

\begin{titlepage}
\begin{flushright}
CECS-PHY-12/04
\end{flushright}
\vspace{2.6cm}
\begin{center}
\renewcommand{\thefootnote}{\fnsymbol{footnote}}
{\LARGE \bf  Gribov copies, BRST-exactness}
\vskip 3mm
{\LARGE \bf and  Chern-Simons theory}
\vskip 20mm
{\large {Marco Astorino$^{1}$\footnote{marco.astorino@gmail.com}, Fabrizio Canfora$^{1},^{2}$\footnote{canfora@cecs.cl}, Jorge Zanelli$^{1},^{2}$\footnote{z@cecs.cl} } }\\
\renewcommand{\thefootnote}{\arabic{footnote}}
\setcounter{footnote}{0}
\vskip 10mm
 \textit{
$^{1}$Centro de Estudios Cient\'{\i}ficos (CECs), Valdivia, Chile   \\
\vskip 1mm
${^2}$Universidad Andr\'es Bello, Av. Rep\'ublica 440, Santiago, Chile} 

\end{center}
\vspace{4 cm}
\begin{center}
{\bf Abstract}
\end{center}
The path integral approach for a 3D Chern-Simons theory is discussed with a focus on the question of metric independence and BRST-exactness in the light of Gribov ambiguity. Copies of the vacuum satisfying the strong boundary conditions and with trivial winding number are shown to exist. This problem is relevant for Yang-Mills theory in 2+1 dimensions as well. Furthermore, it is also shown that there are regular background metrics supporting an infinite number of zero modes of the Abelian Faddeev-Popov operator in the Coulomb gauge. In order to avoid overcounting one could restrict the possible metrics which enter in the gauge fixing but this would imply that, at a quantum level, three-dimensional Chern-Simons theories do depend on the metric.

\end{titlepage}

\section{Introduction}  
The Gribov ambiguity, unveiled in the seminal paper \cite{Gri78} and generalized in \cite{Singer}, states that it is impossible to fix the gauge globally for derivative gauge conditions, which is a serious difficulty for the correct path integral quantization of gauge theories. Here we analyze the Gribov problem in curved three-dimensional spacetimes of different geometries and for different gauge groups. This problem is relevant for any gauge theory in a curved background and, in particular, for those described by Chern-Simons (CS) forms. A three-dimensional CS system is possibly the simplest (non-Abelian) gauge field theory, but it is far from trivial: different systems of physical interest, such as gravitation and high $T_c$ superconductors, among others, are related to CS theories.

Classical CS theories do not require a metric, a feature that can be traced back to their relation with topological invariants of fibre bundles known as characteristic classes \cite{Nakahara}. They are defined in any odd-dimensional spacetime and, at least in 2+1 dimensions, they have no propagating degrees of freedom, which explains why they are commonly referred to as topological field theories (TFT). The standard BRST analysis leads to the conclusion that the CS action is metric-independent not only classically but at the quantum level as well \cite{Birmingham-etal}. In this sense, the discussion of the Gribov problem \cite{Sobreiro-Sorella} could seem to be extremely simple in this case, trivial perhaps. Closer analysis reveals, however, that the problem is far from trivial.\newline

$\bullet $\textbf{\ Relation with Knot theory}

In \cite{guadagnini} it was shown that the perturbative expansion of the partition function of a CS system for SU(2) or SU(N) in the Landau gauge, yields the Jones or (a sub class of) the HOMFLY polynomials, respectively. These results linking CS theory with these two important invariant polynomials of knot theory confirmed the conjecture proposed by Witten \cite{wittenjones}. Following the same approach, but starting from the SO(N) and Sp(N) groups in a variational approach, generates the Kauffman polynomials \cite{astorino}. Moreover, the perturbative expansion in the light-cone gauge yields the Vassiliev invariants \cite{labastida-light, laba-10}, in the form of the Kontsevich integral, which also plays a prominent role in knot theory.

Both in the \textit{light-cone} and in the \textit{temporal} gauge (analyzed in \cite{laba-perez}), a factor must be introduced by hand in order to match the results in the \textit{Landau} gauge or with the non-perturbative approach. The origin of this coefficient --the Kontsevich factor--, which is essential to preserve the topological invariance of Wilson loops' vacuum expectation values, is not fully understood.

From the quantum field theoretical point of view, the difference that results from a change of gauge choice contradicts the expectation that the conclusions in a BRST-invariant theory should not depend on the choice of gauge. Thus, the observed sensitivity of the results on the gauge choice implies either a breaking of BRST symmetry, or a loophole in the link between BRST invariance and gauge independence. A possible explanation for this puzzle might be related to the Gribov ambiguity since, as shown in \cite{Fuj}, the presence of Gribov copies can break BRST symmetry. \newline

\newpage
$\bullet $\textbf{\ The standard hypothesis}

A way to circumvent the Gribov ambiguity in perturbation theory proposed in \cite{AMATI} and advocated, in the context of Chern-Simons theory, by several authors thereafter (see, e.g.,\cite{Adams, Marino}), assumes that the vacuum $A_{\mu }=0$ and its Gribov copies are in different topological sectors, and therefore one cannot be perturbatively accessed from the other.

However, since the early days of TFT it has been widely recognized that  \textit{``it is because there are these Gribov copies that topological field theories are interesting and, in fact, that is the basis of the interpretation of these theories in terms of generalized Morse theory..."} (this quote is taken from the seminal paper \cite{labamorse} where the Morse-theoretical description of TFT was first recognized). So, the existence of Gribov copies on the one hand provides TFT with an interesting and non-trivial structure, and on the other, they severely narrow the very meaning of ``Topological Field Theory" by restricting, as shown in the next sections, the possible background metrics that can be used in practical computations to  ``small neighborhoods" of metrics that do not support Gribov copies. Furthermore, to the best of authors the knowledge, the presence of Gribov copies is one of the few mechanisms which can lead to a spontaneous breaking of BRST symmetry \cite{Fuj} explaining, in this way, the puzzling results described above in Chern-Simons perturbation theory.\newline

$\bullet$\textbf{Statement of the problem}

Classical CS theories are metric-free. Quantum mechanically they are supposed to be metric-independent as well: the expectation value of its energy-momentum tensor is BRST exact, $T^{\mu \nu }=\{Q_{BRST},V^{\mu \nu}\} $, for some $V^{\mu \nu }$. This rests on the assumption that some $V^{\mu \nu }$ exists, and that there are no Gribov ambiguities \cite{Fuj}.

The classical CS equations are also metric-free. However, in order to write down solutions, boundary/initial conditions must be stated and this usually involves coordinates and a metric. In the path integral approach, both the gauge-fixing prescription and the implementation of the Faddeev-Popov trick, require specifying the metric of the space-time where the theory is defined. So, one inevitably requires a metric at some stage.

There is a further issue of metric-dependence associated with the Gribov ambiguity. The Gribov equation --which determines the existence of Gribov copies--, as well as the Faddeev-Popov operator, depend on the space-time metric. Suppose that for a given metric $g_{\mu \nu }^{0}$ there are no Gribov copies in certain domain of the space of connections. Then, there is no guarantee that there will continue to be no copies for metrics obtained by continuous deformations of $g_{\mu \nu }^{0}$. In fact, as the results in \cite{CGO} \cite{ACGO} \cite{CGO2} show in different contexts, the existence of solutions for the Gribov equations depend critically on the spacetime metric. The problem is that the topology of spacetime is not continuous in the metric. For example, an ``infinitesimal amount of cosmological constant" can change the Minkowski space into de Sitter or anti-de Sitter, which have different topologies, different causal structures and asymptotics. Therefore, a slight change in the metric can lead to radical changes in the spectrum of any theory, unless one could restrict the attention to a suitably defined ``small" deformations of the metric $g_{\mu \nu }^{0}$. \newline

\newpage
$\bullet$\textbf{What we do here}

\textbf{1.} Here we analyze the solutions of the Gribov equation, 
\begin{equation}
\nabla ^{i}(U^{\dagger }A_{i}U+U^{\dagger}\partial _{i}U)=0\ , \label{Grib-eq}
\end{equation}
for the Coulomb gauge $\nabla ^{i}A_{i}=0$ and conclude that, in general, it has nontrivial solutions satisfying physically acceptable boundary conditions\footnote{ These commonly known as strong boundary conditions are defined in the next section.} for a large class of space-times, including flat, AdS, black holes and conical singularities, around the vacuum $A=0$. \newline

\textbf{2.} We show that the Faddeev-Popov operator for an Abelian gauge theory in the Coulomb gauge could have a number of normalizable zero modes and that their expressions depend on the spacetime geometry \cite{CGO}. \newline

$\bullet$ \textbf{Consequences}

\textbf{1.} The existence of Gribov copies (as well as of and zero modes of the Faddeev-Popov operator in the Abelian case) around $A=0$ implies that restricting the path integral to \textquotedblleft a region free of ambiguities around the vacuum configuration" --which could be a practical way for setting up a perturbative expansion around the ground state as proposed in \cite{AMATI,Adams,Marino}--, is not a viable option on a generic background. \newline

\textbf{2.} The presence of Gribov copies invalidates the assumptions needed to claim that the expectation value of the energy-momentum tensor is BRST-exact \cite{Fuj}. Therefore, the metric-independence of the quantum CS theory is unwarranted. \newline

$\bullet $ \textbf{Plan of the paper}

In the next section, the Gribov problem on curved spaces is described. Section 3 analyzes the boundary conditions for the copies and the winding number. In Section 4 it is shown that there are infinitely many regular background metrics supporting copies of the vacuum. In the fifth section, the implications for the BRST symmetry are discussed. Eventually, some conclusions are drawn in Section 6.

\section{Gribov problem in curved spaces}  

Consider a gauge theory described by the classical action 
\begin{equation}
I[A]=\int_{M}\mathcal{L}(A,dA)\,,
\end{equation}
where $A$ is a connection in some Lie algebra $\mathcal{G}$ and $M$ is a 2+1-dimensional manifold. Here $\mathcal{L}$ could be a Lagrangian for Yang-Mills or Chern-Simons theories, including gravity. Under the action of the gauge group, the connection $A$ transforms as 
\begin{equation}
A_{\mu }\rightarrow A_{\mu }^{U}=U^{\dagger }A_{\mu }U+U^{\dagger}\partial_{\mu }U\,,  \label{gaugetranformation}
\end{equation}
where $U(x)$ is an element of the (compact) Lie group $G$ with algebra $\mathcal{G}$. The existence of Gribov copies in the Coulomb gauge corresponds the the existence of nontrivial solutions of the system of equations 
\begin{equation}
\nabla ^{i}A_{i}=0\ ,\ \ \nabla ^{i}A_{i}^{U}=\nabla ^{i}(U^{\dagger}A_{i}U+U^{\dagger}\partial _{i}U)=0\,,
\end{equation}
($i$ being the spatial indices) subject to the boundary condition that $U(x)\rightarrow C_{G}$ sufficiently fast for $|x|\rightarrow \infty $, where $C_{G}$ is an element of the center of the gauge group\footnote{The center of the gauge group is defined as the set of elements which commute with all the element of $G$.} $G$, (the boundary conditions will be
discussed in the next section).

Let us consider a three-dimensional Euclidean metric of the form 
\begin{eqnarray}
ds^{2} &=&h^{2}(r)d\tau ^{2}+f^{-2}(r)dr^{2}+r^{2}d\phi ^{2}\ , \label{metrcoulomb} \\
0 &\leq &r\leq \infty \ ,\ \ \ 0\leq \tau \leq 2\pi \ ,\ \ \ 0\leq \phi \leq 2\pi \ ,  \notag
\end{eqnarray}
where $\tau $ is the Euclidean time, which we assume to be compactified so that the topology of the manifold is $S_{1}\times \Sigma _{2}$. In the cases in which $h(r)=f(r)$, the above geometries may correspond, for instance, to the Euclidean rotation of a static, circularly symmetric space-time, including: flat space, $f^{2}=1$; (A)dS, $f^{2}=1(\pm)r^{2}/l^{2}$; point particles (conical defects), $f^{2}=M+r^{2}/l^{2}$, \cite{Deser-Jackiw-tHooft}; static black holes, $f^{2}=-M+r^{2}/l^{2}$, \cite{BTZ}. Clearly, these metrics are continuously deformable among themselves in any local patch, but they define quite different global geometries.

The spatial slices defining the Coulomb gauge are the $\tau =constant$ surfaces, with induced metric 
\begin{equation}
ds^{2}=f^{-2}(r)dr^{2}+r^{2}d\phi ^{2}= g_{ij}dx^{i}dx^{j}\ ,\ \ i,j=1,2\ \ ,  \label{spatialslice}
\end{equation}
The Coulomb gauge condition is 
\begin{equation}
A_{\tau }^{a}=0\ ,\ \ \ \nabla ^{i}A_{i}^{a}=\frac{1}{\sqrt{\det g}}\partial_{j}\left( \sqrt{\det g}g^{ji}A_{i}^{a}\right) =0,\ .  \label{2Dcoulombgauge}
\end{equation}
In order to fix ideas, let us take the case $G=SU(2)$ , with a gauge transformation of a form analogous to the Gribov ansatz in 3+1 dimensions, 
\begin{equation}
U(x)=\exp \left( i\frac{\alpha (r)}{2}\overrightarrow{n}\cdot \overrightarrow{\sigma }\right) \ ,\ \ \overrightarrow{n}\cdot \overrightarrow{\sigma }=n^{1}\sigma _{1}+n^{2}\sigma _{2}\,, \label{2Dcoulombansatz}
\end{equation}
where $\overrightarrow{n}=(\cos \phi ,\sin \phi )$ is the two-dimensional unit radial vector and $\sigma _{1}$, $\sigma _{2}$ are the standard Pauli matrices. It can be seen that $\left( \overrightarrow{n}\cdot \overrightarrow{\sigma }\right) ^{2}=\mathbf{1}_{2\times 2}$, and $U^{\dag}=U^{-1}$.

As it is well known, in the path integral formalism, an ambiguity in the gauge fixing may lead to a smooth zero mode of the Faddeev-Popov (FP) operator satisfying suitable boundary conditions. In order to define the path integral in the presence of Gribov copies, it has been suggested to exclude classical $A_{\mu }$ backgrounds which generate zero modes of the FP operator (see, in particular, \cite{Gri78} \cite{Zw82} \cite{Zw89} \cite{Zw93} \cite{DZ89}\ \cite{Zwa96} \cite{Va92}). This possibility is consistent with the usual perturbative point of view on flat space-times since, in the case of $SU(N)$ Yang-Mills theories, for a``small enough" potential $A_{\mu }$ (with respect to a suitable functional norm \cite{Va92}), there are no zero-modes of the Faddeev-Popov operator in the Landau or Coulomb gauge\footnote{The issue of gauge fixing ambiguities cannot be ignored since even if gauge fixing choices free of Gribov ambiguities can always be found, still the presence of Gribov ambiguities in other gauges gives rise to a breaking of the BRST symmetry at a non-perturbative level (see, for instance, \cite{Fuj}).}. Indeed, the worst possible pathology would be to have a copy of the vacuum $A_{\mu }=0$ fulfilling the strong boundary conditions (defined in the next section) since, in this case, not even usual perturbation theory leading to the standard Feynman rules in the Landau or Coulomb gauge would be well defined.

For a nonabelian theory in flat ($3+1$)-dimensional space, a perturbative region around $A_{\mu }=0$ free of copies does exist but. However, as shown in \cite{CGO} \cite{ACGO} \cite{CGO2}, the situation becomes more delicate in a curved background since on static asymptotically AdS spacetimes even the trivial vacuum $A_{\mu }=0$ could have Gribov copies. These results suggest to study how the gauge fixing process and the appearance of copies of the vacuum in CS theories change when the background metric is deformed. The existence of copies of the vacuum is determined by the non-trivial solutions of the Coulomb gauge condition, 
\begin{equation}
\nabla ^{i}\left( U^{-1}\partial _{i}U\right) =0\   \label{2DCoulomb}
\end{equation}
In the ansatz (\ref{metrcoulomb}) this equation reduces to 
\begin{equation}
rf\frac{d}{dr}\left( rf\frac{d\alpha }{dr}\right) =\sin \alpha \ . \label{2Dpendulum}
\end{equation}
It is useful to introduce a new variable $s$ such that 
\begin{equation}
s=\int^r \frac{dr'}{r'f(r')} + s_0, \label{2Dchange}
\end{equation}
so that in this new variable Eq. (\ref{2Dpendulum}) becomes the static sine-Gordon equation, 
\begin{equation}
\frac{d^{2}\alpha }{ds^{2}}=\sin \alpha \, .    \label{sinegordon}
\end{equation}
A first integral of this equation is 
\begin{equation}
E=\frac{1}{2}\left( \frac{d\alpha }{ds}\right) ^{2}+\cos \alpha \ ,
\end{equation}
where $E$ is an integration constant, and the formal solution of Eq.(\ref{sinegordon}) reads 
\begin{equation}
s-s_{0}=\pm \int^{\alpha }\frac{dx}{\sqrt{2\left( E-\cos x\right) }}\ , \label{s(alpha)}
\end{equation}
where $s_{0}$ is another integration constant. As shown in the next sections, there is always at least one choice of $E$ which gives rise to a non-trivial solution satisfying physically sensible boundary conditions (described below). Moreover, we will show that there are infinitely many regular background metrics supporting copies of the vacuum with trivial winding number.

\section{Boundary conditions}  

Let us now discuss the boundary conditions for the function $\alpha$ in the Schwarzschild-like coordinates of Eq. (\ref{metrcoulomb}). Not every regular smooth solution, $U(x)$, of the Gribov equation (\ref{2DCoulomb}) represents a physically acceptable gauge transformation. A \textit{proper gauge transformation}, is an everywhere regular gauge transformation which does not change the definition of the (non-Abelian) charges as surface integrals at spatial infinity. This requirement defines the so-called \textit{strong boundary conditions}; a smooth gauge transformation which does not satisfy this property can be discarded (see, e.g., \cite{Benguria:1976in}, \cite{HRT},  \cite{Henneaux-Teitelboim}, \cite{Gri78}). The strong boundary conditions for the function $\alpha $ in these coordinates are
\begin{align}
&\alpha \underset{r\rightarrow \infty }{\rightarrow }2n\pi +O(1/r^{\eta }) ,\ \ \eta >0\ ,  \label{adsr-strong} \\
&\alpha \underset{r\rightarrow 0}{\rightarrow }2m\pi +O(r^{\gamma })\ ,\ \quad \gamma >0\ ,\  m,n \in \mathbb{Z}. \label{origin}
\end{align}
The first condition ensures that the group element of the form $U\left( x^{\mu }\right) =\exp \left( i\frac{\alpha \left(r\right) }{2}x^{i}\sigma _{i}\right)$ approaches $\pm I$ at infinity and, consequently, it does not change the charges by modifying the fields at spatial infinity. Condition (\ref{origin}), in turn, must hold in order for the copies to be regular at the origin.

The analysis of Eq. (\ref{2Dpendulum}) is more transparent in terms of the coordinate $s$ as in Eq. (\ref{s(alpha)}). We consider different $f$s corresponding to some 2+1 dimensional geometries, including the AdS space-time, a conical defect, the BTZ black hole and flat space (to be discussed in the next section). The range of the coordinate $s(r)$ defined in Eq. (\ref{2Dchange}) corresponding to the these cases is presented in the
following table, 
\begin{equation}
\begin{array}{|l|c|c|c|}
\hline
Spatial\ \ Geometry\  & f(r) & r-range & s-range \\ \hline
Flat & 1 & \left[ 0,\infty \right) & \left( -\infty ,+\infty \right) \\ 
AdS & \sqrt{1+\frac{r^{2}}{l^{2}}} & \left[ 0,\infty \right) & \left(\infty , 0 \right) \\ 
Conical\ defect & \sqrt{M+\frac{r^{2}}{l^{2}}} & \left( 0,\infty \right)&  \left(\infty , 0 \right) \\ 
Black\ hole & \sqrt{-M+\frac{r^{2}}{l^{2}}} & \left( l\sqrt{M},\infty \right) & \left( 0,\frac{\pi }{2\sqrt{M}}\right) \\ 
\hline
\end{array}
\ \   \label{table1}
\end{equation}
In the case of the conical defect and the black hole, the freedom to choose the integration constant ($c$) was used to adjust the range of $s$. It can be observed that in the case of AdS as well as for the conical defect, the range in $s$ is reversed with respect to the range in $r$ in the sense that $s(r\rightarrow \infty )=0$ and $s\left(r\rightarrow 0\right) =\infty$. 

Gribov's equation (\ref{2Dpendulum}) depends on the spacetime geometry, even if in terms of the coordinate $s$ (\ref{sinegordon}) looks the same for all the metrics listed above. The difference manifests itself in the original radial coordinate $r$. While for the cases of AdS and conical defect, the range of $s$ is the half-line, the range of $s$ for the black hole is necessarily bounded.

Both for AdS and for the conical defect, it is easy to see that the strong boundary conditions (\ref{adsr-strong}) can be satisfied for certain values of the integration constants. In particular, for $E=1$, $\alpha$ can be expressed in terms of elementary functions, 
\begin{equation}
\alpha (s,E=1)=4\arccos [\tanh (s-s_{0})]\ .  \label{E=1}
\end{equation}
Thus, irrespectively of the value of $s_{0}$, $\alpha \rightarrow 0$ for $s\rightarrow \infty$, satisfying the strong boundary conditions (\ref{origin}). The constant $s_{0}$ must be set to zero in order to comply with condition (\ref{adsr-strong}).

In the black hole case, $E\leq 1$ would lead to an unbounded range for the coordinate $s$, so the solution only exists for $E>1$. It can be easily seen that in this case there are three integration constants ($E$, $s_0$ and $c_2$) subject to the two requirements (\ref{adsr-strong}) and (\ref{origin}), leading to a one-parameter family of solutions.

\section{Background geometries supporting copies of the vacuum}  

We now show that there are infinitely many regular background metrics supporting copies of the vacuum satisfying the strong boundary conditions (\ref{adsr-strong}, \ref{origin}). Here we follow the idea of Ref. \cite{Henyey} to obtain the background field (which in the present case is the metric) from Gribov's equation for the copies.\\

\subsection{Copies of vacuum in flat space}  

It is often assumed that a given configuration $A$ and its Gribov copies $A^{U}$ have different winding numbers. In fact, for the Coulomb gauge in 2+1 dimensions, one requires that $A_{i}=U^{-1}\partial _{i}U$, with 
$U(x)=\exp (i\frac{\alpha (r)}{2} \vec{n} \cdot \vec{\sigma})$, and that $\alpha$ satisfies the condition (\ref{adsr-strong}). Following the usual arguments, the above boundary conditions define a mapping from the spatial sphere at infinity (which in this case is $S^{1}$) to the gauge group $SU(2)$. As is well known, the winding number must vanish because the corresponding homotopy group, $\pi _{1}(SU(2))$, is trivial \cite{Nakahara} (all loops in $SU(2)$ are contractible). This argument holds, in particular, in the flat case, $f(r)=1=h(r)$. 

However, one can still argue that the copies which have been constructed here in the $SU(2)$ case cannot be deformed continuously to the identity of the gauge group since they satisfy the two requirements (\ref{adsr-strong}) and (\ref{origin}) with $m-n=\pm1$. Therefore, such copies interpolate between the $I$ at the origin and $-I$ at infinity, or vice-versa. For this reason, these configurations possess a topological charge related to the sine-Gordon soliton number, which is independent from the winding number of the the gauge transformation.

\subsection{Copies of vacuum in other geometries}  

We now address the question whether copies of the vacuum with trivial winding number can exist in other geometries. For this purpose, it is convenient to interpret (\ref{2Dpendulum}) as an equation for the metric coefficient $f(r)$, in terms of $\alpha $. A first integral of (\ref{sinegordon}) gives 
\begin{equation}
f^{2}(r)=\frac{C_{1}-2\cos \alpha }{\left( r\alpha ^{\prime }\right) ^{2}}\ , \label{f(alpha)}
\end{equation}
where $C_{1}=2E$ is the integration constant. The above equation is a suitable starting point to analyze the question of what kind of metrics support copies of the vacuum in the Coulomb gauge. We assume that $\alpha $ is everywhere regular and smooth and satisfies the boundary conditions 
\begin{equation}
\alpha (r)\underset{r\rightarrow 0}{\approx }\alpha _{0}r+o(r^{2})\ ,\ \
\alpha (r)\underset{r\rightarrow \infty }{\approx }2\pi +o(r^{-1}).
\label{req0}
\end{equation}
Regularity of the metric at the origin requires that $f^{2}$ should behave as 
\begin{equation}
f^{-2}\left( r\right) \underset{r\rightarrow 0}{\approx }\ 1+kr^{2}+o(r^{3})\,  \label{req1}
\end{equation}
where $k$ is a constant. On the other hand for $r\rightarrow \infty $, $f^{2} $ should behave as 
\begin{equation}
f^{-2}(r)\underset{r\rightarrow \infty }{\approx }\ 1+\frac{k_{1}}{r} +o(r^{-2})\ ,  \label{req1.25}
\end{equation}
for asymptotically flat spacetimes, or as 
\begin{equation}
f^{-2}(r)\underset{r\rightarrow \infty }{\approx }\ \frac{k_{2}}{r^{2}} +o(r^{-3})\ ,  \label{req1.5}
\end{equation}
for asymptotically AdS spacetimes. It is easy to see that this implies that the denominator of Eq. (\ref{f(alpha)}) must vanish for $r\rightarrow 0$, which requires choosing $C_{1}=2$, so that Eq. (\ref{f(alpha)}) must read 
\begin{equation}
f^{2}(r)=\frac{2(1-\cos \alpha )}{\left( r\alpha ^{\prime }\right) ^{2}}\ \ . \label{inverscattcoulomb2}
\end{equation}
This equation shows that there are clearly infinitely many metrics of the form (\ref{metrcoulomb}) supporting copies of the vacuum that satisfy the  strong boundary conditions. Indeed, for every smooth and bounded function $\alpha (r)$ there is one function $f(r)$ satisfying (\ref{inverscattcoulomb2}). Since there are infinitely many $\alpha $'s, monotonically increasing\footnote{We take this restriction in order to avoid the possible occurrence of horizons.} from $\alpha (0)\ =0\ \ \text{to}\ \ \alpha (\infty )\ =2\pi $, satisfying the strong boundary conditions (\ref{req0}), there are also infinitely many metrics that support copies of the vacuum. In particular, for $\alpha (r)=4\arctan r$, (\ref{f(alpha)}) yields $f^{2}(r)=1$. Thus, among the infinitely many background metrics of the form (\ref{metrcoulomb}) that support copies of the vacuum in the Coulomb gauge, there is also the flat Euclidean metric\footnote{Since the metric component $g_{00}$ does not enter in Gribov's equation for the Coulomb gauge, one can take $h(r)=1$. When $h=1$ the solutions presented here as copies in the Coulomb gauge can also be interpreted as copies in the Landau gauge.}.

\subsection{Zero modes of the Abelian Faddeev-Popov operator}   

On a curved background, even an Abelian gauge theory could inherit gauge-fixing pathologies. In particular, following the steps of \cite{CGO}, it can be shown that there are regular background metrics that support an infinite number of smooth and normalizable zero-modes of the Faddeev-Popov operator in the Coulomb gauge. This pathology is more serious than the appearance of Gribov copies of the vacuum in the non-Abelian case (where it is an effect of the non-linearity of the gauge transformation), since it leads to the vanishing of the Faddeev-Popov determinant. Thus, even the simplest gauge systems, like the Abelian 3D CS theory or QED, may suffer from gauge-fixing issues.

The equation for the zero modes of the Faddeev-Popov operator in the Coulomb gauge is 
\begin{equation}
\nabla^i\nabla_i\psi =0\ ,  \label{vaczeromodes}
\end{equation}
where the gauge transformation is $U=\exp(i\psi)$ and $\psi(r,\phi)$ is a scalar function. Here $\nabla_i$ is the Levi-Civita covariant derivative corresponding to the background metric along the spatial directions.

The existence of smooth normalizable solutions of Eq. (\ref{vaczeromodes}) represents a direct obstruction to define a perturbation theory. In this case, the Faddeev-Popov operator would vanish identically since the Laplacian contains zero modes and the perturbative procedure breaks down.

Consider a three-dimensional static spherically symmetric metric of the form (\ref{metrcoulomb}), and let us decompose the gauge parameter $\psi $ as 
\begin{equation}
\psi =F(r)\sin (m\phi )\ ,  \label{separation}
\end{equation}
$m$ being an integer number. Using Eq. (\ref{separation}) one can see that (\ref{vaczeromodes}) reduces to 
\begin{equation}
\left( rf\right) ^{2}F''+rf(rf)' F'^2 F=0\ ,  \label{sessea}
\end{equation}
which can also be written as 
\begin{equation}
\frac{d^{2}}{ds^{2}}F-m^{2}F=0\,.
\end{equation}
The solutions are $F(s)=F_{0}^{(+)}\exp (ms)+F_{0}^{(-)}\exp (-ms)$. For the black hole case $s$ is bounded and both solutions are admissible; for the other cases, $F$ remains bounded only if $F_{0}^{(+)}=0$.

It is customary to require the gauge transformation to be normalizable in function space, for instance as \cite{Va92} 
\begin{equation}
\mathcal{N}(\psi) :=\int_{\Sigma }\sqrt{\det g}d^2\Sigma \left( \nabla_i \psi \right) \left( \nabla^i \psi \right) <\infty .
\label{normcond2}
\end{equation}
For the solution $\psi=F^{(-)}_0 \exp(-ms)\sin(m\phi)$, this integral is 
\begin{equation}
2\pi m^2 \int_0^{\infty} ds F^2=2\pi m^2 \left(F_0^{(-)}\right)^2.
\end{equation}

So, in all the geometries discussed in the table (\ref{table1}) except for the flat case, there exist a family of zero modes $\psi $ of the form (\ref{separation}), parametrized by the integer $m$ and the continuous real parameter $F_{0}^{(-)}$. In all these cases, the Abelian gauge transformation $U$ approaches a phase for $s=0$ and the identity for $s\rightarrow \infty $. In the flat case, the solutions are not normalizable, and according to the common practice they would be ruled out.

After performing the Wick rotation $t\rightarrow i\tau $ in the above metric, one can easily construct examples of smooth compact manifolds supporting zero modes of the vacuum Faddeev-Popov operator for the Landau gauge. This implies, for example, that computing the Gauss linking number in the Landau gauge, as is commonly done in flat spacetime, could be troublesome. In particular, difficulties should be expected in the perturbative expansion of the Wilson loop around the vacuum of an Abelian gauge theory, due to the vanishing of the Faddeev-Popov operator. This example shows that in curved spaces a careful interpretation for the linking number as the vacuum expectation value of the Wilson loop is needed.

\section{Metric independence and BRST exactness of CS theories}  

Let us now briefly review the usual formal definition of TFT (for a detailed review, see \cite{Birmingham-etal}). The key idea is to extend the concept of a field theory which is \textit{metric independent} to a quantum level. In this way one expects that such a theory could properly capture topological features of the manifold on which it is defined.

In order to fix ideas, let us consider the Chern-Simons theory for the $SU(2)$ group in the Landau gauge. The total action $S_{T}$, which includes ghosts and gauge fixing term is 
\begin{align*}
S_{T}& =S_{(0)}+S_{g.f.}+S_{ghosts}\ , \\
S_{(0)}& =\frac{k}{4\pi }\int A\wedge dA+\frac{2}{3}A\wedge A\wedge A\ , \\
S_{g.f.}& =\int g^{\mu \nu }b^{a}\partial _{\mu }A_{\nu }^{a}\ , \\
S_{ghosts}& =\int g^{\mu \nu }\overline{c}^{a}\partial _{\mu }\nabla_{\nu}c^{a}\ ,
\end{align*}
where $\nabla _{\nu}$ is the covariant derivative, $c^a$ and $\overline{c}^a$\ are the ghost and anti-ghost fields and $b^a$ is the Lagrange multiplier enforcing the Landau gauge condition. The classical CS action $S_{(0)}$ does not depend on the metric so $g^{\mu \nu }$ only enters in the gauge fixing and ghosts terms, which are formally BRST-exact. The energy momentum tensor $T_{\mu \nu }$ is defined as the variational derivative of the total action with respect to the metric 
\begin{equation}
S_{T}\left( g^{\mu \nu }+\delta g^{\mu \nu }\right) -S_{T}\left( g^{\mu \nu}\right) \ \overset{def}{=}\ \frac{1}{2}\int d^{3}x\sqrt{g}\delta g^{\mu \nu }T_{\mu \nu }\ .  \label{tmunu1}
\end{equation}
Since the metric only appears in the BRST-exact terms $S_{g.f.}$ and $S_{ghosts}$, their variations are also BRST exact. The energy momentum tensor can then be written as 
\begin{equation}
T_{\mu \nu }=\left\{ Q_{BRST},V_{\mu \nu }\right\} \ ,  \label{tmunu2}
\end{equation}
where $V_{\mu \nu }$ is some local function of the fields and the metric, whose explicit expression is not needed here. From Eqs. (\ref{tmunu1}) and (\ref{tmunu2}) one concludes that the variation of the partition function $Z$
with respect to the metric corresponds to the vacuum expectation value of the energy-momentum tensor: 
\begin{equation}
\frac{\delta }{\delta g^{\mu \nu }}Z =i\int DA_{\mu }^{a}Dc^{a}D\overline{c}^{a}Db^{a}\exp \left( iS_{T}\right) \left\{Q_{BRST},V_{\mu \nu}\right\} \, .
\label{tmunu3}
\end{equation}
When the BRST symmetry is unbroken, Eq. (\ref{tmunu3}) implies the metric independence of the CS theory since the vacuum expectation values of BRST-variations vanish.

This standard argument, however, may fail in the presence of Gribov copies which break BRST symmetry \cite{Fuj}. Thus, even if one assumes that for a given metric $g_{\mu \nu }^{(0)}$ no Gribov copies appear, this may not continue to be true upon variation of the metric, as required to compute the energy momentum tensor. In particular, as we have shown in the previous sections, $g_{\mu \nu}^{(0)}+\delta g^{\mu \nu}$ may support Gribov copies. In these cases the partition function would not be differentiable anymore with respect to the metric, unless one could restrict the theory to a ``suitable neighborhood" of $g_{\mu \nu }^{(0)}$ in the functional space of metrics, which could be free of gauge-fixing ambiguities. 

\section{Discussion and perspectives}   

We have presented examples of smooth 3-dimensional Euclidean background geometries (including flat space) supporting Gribov copies of the vacuum in the Coulomb gauge satisfying physically reasonable (e. g., ``strong") boundary conditions and with trivial winding number. We have also shown that it is possible to construct regular geometries supporting an infinite number of smooth normalizable zero-modes of the Faddeev-Popov operator in the Coulomb gauge for an Abelian gauge theory. In all of these cases it is observed that the existence of the Gribov copies depends critically on the metric of the background spacetime.

As discussed in the introduction in connection with knot theory, perturbative computations in different gauge fixings apparently lead to inequivalent results, something one would not expect from a BRST-invariant theory. The results discussed here could be closely related to this issue through the work of Fujikawa \cite{Fuj}, who showed that the appearance of Gribov copies in some gauge could break BRST symmetry of the theory at the non perturbative level.

An interesting possibility would be to restrict the class of allowed background metrics for a given topological field theory. Hence, the concept of a topological field theory should be valid only in a neighborhood of a suitable metric which is free of gauge fixing pathologies. This point of view is reminiscent of the Gribov-Zwanziger approach to QCD. However, in the present case the restriction should not be in the domain of integration of the path integral but on the allowed background metrics. 

The discussion in the previous section shows that such neighborhood may not exist. In this case, the perturbative regime of the topological theory itself may not be well defined through the path integral. 

\section*{Acknowledgments}

This work has been partially supported by FONDECYT through grants 1120352, 1100755, 1085322, 1100328, 1110102, and 3120236, and by the ``Southern Theoretical Physics Laboratory" ACT-91 grant from CONICYT. The Centro de Estudios Cient\'{\i}ficos (CECs) is funded by the Chilean Government through the Centers of Excellence Base Financing Program of CONICYT. F. C. is also supported by Proyecto de Inserci\'on CONICYT 79090034.


\end{document}